\documentclass[aps,pre,reprint,groupedaddress]{revtex4-1}

\usepackage{graphicx}
\usepackage{amsmath,bm,dsfont}
\usepackage{siunitx}
\usepackage{amssymb}
\usepackage{mathrsfs}
\usepackage{color}
\usepackage{times}

\newcommand{\kt}{k_\mathrm{B}T}
\newcommand{\Sum}{\sum_{Q,\bm q}}
\newcommand{\hq}{h_{Q,\bm q}}
\newcommand{\Jq}{J_{Q,\bm q}}
\newcommand{\hmq}{h_{-Q,-\bm q}}
\newcommand{\Jmq}{J_{-Q,-\bm q}}
\newcommand{\Smq}{S_{-Q,-\bm q}}
\newcommand{\Sq}{S_{Q,\bm q}}
\newcommand{\Aq}{A(Q,q)}

\newcommand{\BesselJ}{{\mathop{\mathrm{J}}}}
\newcommand{\BesselK}{{\mathop{\mathrm{K}}}}

\begin{document}


\title{Interaction and self-assembly of membrane-binding\\and membrane-excluding colloids embedded in lamellar phases}

\author{Ruben Zakine\textit{$^{a}$}, Dasith de Silva Edirimuni\textit{$^{a}$}, Doru Constantin\textit{$^{b}$}, Paolo Galatola\textit{$^{a}$} and Jean-Baptiste Fournier$^{\ast}$\textit{$^{a}$}}

\affiliation{$^{a}$Laboratoire``Mati\`ere et Syst\`emes Complexes" (MSC), UMR 7057 CNRS, Universit\'e Paris 7 Diderot, 75205 Paris Cedex 13, France.\\
$^{b}$Laboratoire de Physique des Solides, CNRS, Universit\'e Paris-Sud,
Universit\'{e} Paris-Saclay, 91405 Orsay Cedex, France.}

\date{\today}

\begin{abstract}
Within the framework of a discrete Gaussian model, we present
analytical results for the interaction induced by a lamellar phase
between small embedded colloids. We consider the two limits of particles
strongly adherent to the adjacent membranes and of particles
impenetrable to the membranes. Our approach takes into account the
finite size of the colloids, the discrete nature of the layers, and
includes the Casimir-like effect of fluctuations, which is very
important for dilute phases. Monte Carlo simulations of the
statistical behavior of the membrane-interacting colloids account semi-quantitatively,
without any adjustable parameters, for the experimental data measured on
silica nanospheres inserted within lyotropic smectics. We predict the existence
of finite-size and densely packed particle aggregates originating from the competition between attractive
interactions between colloids in the same layer and repulsion
between colloids one layer apart.
\end{abstract}

\maketitle 

\section{Introduction}\label{sec:intro}

Assembling nanoparticles into organized superstructures is one of the
most important topics in contemporary materials science. The sought-for
organization concerns positional order, since the properties of the
individual particles can be tuned via coupling to their neighbors,
yielding enhanced or even completely novel characteristics (as for
metamaterials).  However, the orientational order is equally important:
in the case of anisotropic nanoparticles, aligning them is an essential
step towards propagating this anisotropy to the macroscopic scale of the
final material. Other important requirements concern the finite size of
the resulting assemblies and their shape (e.g. elongated, flat, isometric etc.).

In this context, using liquid crystal matrices, which are both ordered
and soft (and thus easily processable) is a promising design option for
hybrid materials with original properties~\cite{Bisoyi:2011}. Many groups
have employed this strategy, using either nematic phases (which only
exhibit orientational order)~\cite{Pratibha:2009,Liu:2014} or smectic
phases (with an additional positional
order)~\cite{Wojcik:2009,Coursault:2012,Lewandowski:2013}. It is clear
that the fundamental problem of the interaction between inclusions in
liquid crystals can have very practical applications.

The examples above concern thermotropic phases (consisting of a single
type of molecules). Another large class of systems is that of lyotropic
phases, where self-assembled entities (micelles, bilayers, etc.) are
dispersed in a (usually aqueous) solvent.

Lyotropic liquid crystals are particularly suitable for controlling the
self-assembly process, because their elastic moduli are lower than those
of thermotropics, so that the induced interactions can be weaker
(comparable to $k_B T$) and because their properties (spacing,
flexibility, electrical charge etc.) are easily tuned via the
composition~\cite{Liu:2010,Venugopal:2011}. In addition, their
intrinsically heterogeneous nature (bilayers of amphiphilic molecules
separated by water layers) opens the possibility of confining within the
same phase particles with different chemical
affinities~\cite{Wang:1999,Firestone:2001}.

For all these reasons, particle inclusions in lyotropic phases have
raised the interest of numerous groups all over the
world~\cite{Constantin:2014}. Both the structuring effect of the host
lamellar phase on the inclusions~\cite{Henry:2011} and the influence
of the latter on the dispersing phase~\cite{Yamamoto:2005} have been studied.

The distinctive features of lyotropic phases must also be accounted for
in a realistic model for the induced interactions: i) The boundary
conditions must be stated in terms of particle interaction with discrete
(and generally impenetrable) bilayers rather than with a continuous
director field. ii) These soft systems exhibit strong fluctuations, so
the model should take into account fluctuation-induced, Casimir-like
interactions.

The model presented here deals with lamellar lyotropic phases and
includes both these aspects. Using a discrete Gaussian
description~\cite{Holyst91,Lei95,safran_book}, we treat the coupling
between the lamellae and the colloids in an almost exact manner. The
fluctuations are accounted for rigorously, since the interaction results
from an integral over all membrane configurations. In terms of boundary
conditions, we consider two types of colloids: membrane-binding, which
adhere strongly to the neighboring bilayers, and membrane-excluding,
which exhibit hard-core repulsion with respect to the membranes. The
adhesion and the repulsion are considered very strong with respect to
the other energy scales in the problem, so the colloid is simply
described by its diameter. In both cases the membrane-colloid
coupling is nonlinear. Interestingly, for membrane-excluding colloids
smaller than the membrane separation, the situation is analogous to the
Casimir effect, since the presence of the colloids merely suppresses
fluctuation modes of the lamellar phase. 

Our paper is organized as follows. In Section~\ref{sec:lamellar}, we
describe our model of the lamellar
phase~\cite{Holyst91,Lei95,safran_book} and we describe the lamellae
fluctuations. In Section~\ref{sec:binding}, we study the interactions
between membrane-binding colloids and the deformations they create in
the membrane. In Section~\ref{sec:excluding}, we study the interactions
between membrane-excluding colloids and the deformations they create in
the membrane. In Section~\ref{sec:MC} we investigate the equilibrium statistical
behavior of the interacting colloids and their self-assembly by Monte Carlo
simulations. In Section~\ref{sec:saxs}, we compare our simulations to the
experimental results of ref.~\citenum{Beneut08}. Finally, in Section~\ref{sec:conc}
we resume our results and we conclude.

\section{Lamellar phase model}\label{sec:lamellar}

Lamellar phases are stacks of membranes in an aqueous environment.
Membranes interact through different mechanisms: attractive van der
Waals forces, repulsive hydration forces at very short separations,
screened electrostatic forces, and the so-called Helfrich long-range
repulsion arising from the loss of entropy associated with the
confinement of the transverse membrane fluctuations~\cite{Helfrich78}.
The latter may dominate for membranes with weak bending rigidities,
i.e., typically for surfactant systems~\cite{Roux88}. There may also be
attractive fluctuation--induced interactions originating from counterion
correlations~\cite{Jho10}.

Lipids form lamellar phases of bilayer membranes where the layer spacing
$d$ is usually comparable to the membrane thickness $\delta$, i.e., a
few \si{\nano\meter}. Conversely, surfactant systems can form lamellar
phases with $d\gg\delta$, and even unbound lamellar phases. An exact
theoretical modelization of the elasticity of lamellar phases, taking
into account all the different interactions, is very difficult to
achieve. We therefore limit ourselves to the discrete Gaussian elastic
theory of lamellar phase~\cite{Holyst91,Lei95,safran_book}, as described
below.

\begin{figure}[t]
\centerline{\includegraphics[width=\columnwidth]{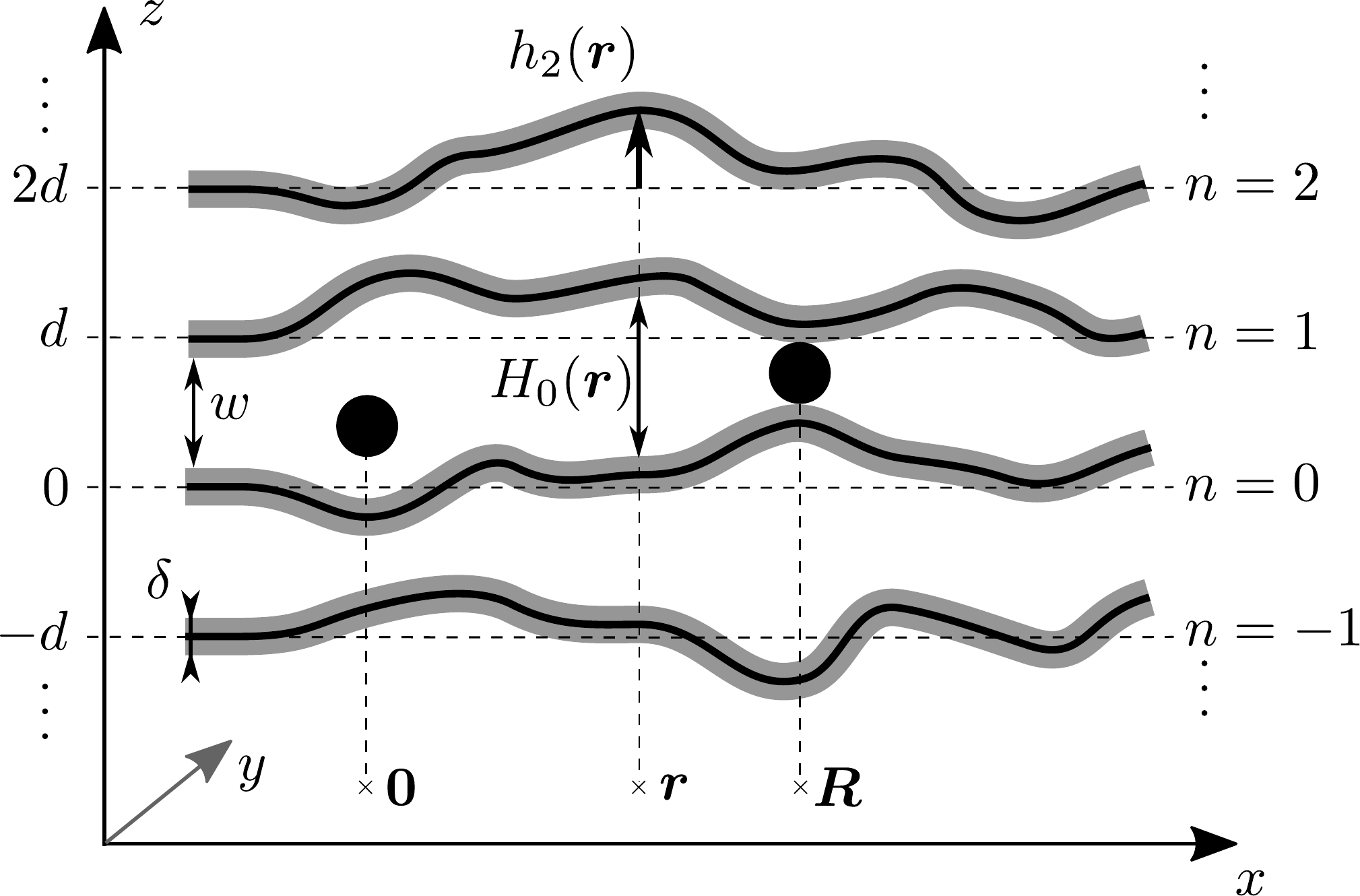}}
\caption{Parametrization of the lamellar phase (cross section). The membranes, of
thickness $\delta$, are drawn in gray and their midsurfaces are
represented as black lines. The average lamellar spacing and water
thickness are $d$ and $w$, respectively. The layer displacements are
described by the functions $h_n(\bm r)$ and the gap between the layers
by the functions $H_n(\bm r)$, as indicated. Two colloids are
represented as black disks: the one on the left is a membrane-excluding
colloid having only excluded volume interactions with the membranes, the
one on the right is a membrane-binding colloid that sticks to the
membranes.}
\label{schema}
\end{figure}

We consider a lamellar phase consisting of $N$ parallel membranes of
thickness $\delta$ (see Fig.~\ref{schema}). We assume that in the
homogeneous, equilibrium state, the thickness of the water layers
between the membranes is $w$ and the repeat distance is $d$, with
\begin{align} d=w+\delta.  \end{align} In a Cartesian reference frame
$(\bm r,z)\equiv(x,y,z)$, we parametrize  the shape of the $n$-th
membrane by the height function $z_n(\bm r)=nd+h_n(\bm r)$, which
represents its elevation above the plane  $z=0$. In addition to the
bending energy of each membrane~\cite{Helfrich73}, the elastic energy
includes the most general interaction that is quadratic in the $h_n$'s,
couples only adjacent membranes, and complies with global translational
invariance~\cite{Holyst91,Lei95,safran_book}:
\begin{align}
\label{eq:H}
\mathcal{H}=\sum_{n=0}^{N-1}\int\!d^2r\left[
\frac{\kappa}{2}\left(\nabla^2h_n\right)^2
+\frac{B}{2}\left(h_{n+1}-h_n\right)^2 \right].
\end{align}
Here, $\kappa$ is the bending stiffness of the membranes and $B$ is an
effective compression modulus, which accounts for all the interactions
between the layers. Note that we have assumed that the membrane
undulations are gentle enough so that the Gaussian approximation of the
curvature energy can be used~\cite{Helfrich73}. The bulk moduli for
layer compression and layer curvature~\cite{degennes69} are $B_3=Bd$ and
$K=\kappa/d$, respectively.  

It is convenient to work with dimensionless quantities. We use $\kt$ to
normalize the energies, $\xi=(\kappa/B)^{1/4}$ to normalize the lengths
parallel to $(x,y)$, and $\xi_\parallel=\xi\sqrt{\kt/\kappa}$ to
normalize $h_n$ and all the lengths parallel to $z$, including $d$,
$\delta$, $w$ and the colloid diameters, hereafter called $a$ and $b$.
The characteristic lengths
$\xi$ and $\xi_\parallel$ are linked to the de Gennes penetration length
$\lambda$~\cite{degennes69} and to the Caill\'e exponent
$\eta$~\cite{Caille72} by the relations $\xi=\sqrt{\lambda d}$ and
$\xi_\parallel=d\sqrt{2\eta/\pi}$.  From now on, unless otherwise
specified, all quantities will be in dimensionless form. Thus, the
Hamiltonian~(\ref{eq:H}) becomes
\begin{align}
\mathcal{H}&=\sum_{n=0}^{N-1}\int\!d^2 r\left[
\frac{1}{2}\left(\nabla^2 h_n\right)^2
+\frac{1}{2}\left(h_{n+1}-h_n\right)^2
\right].
\end{align}

It is also convenient to work in Fourier space. We thus define
\begin{align}\label{hn}
h_n(\bm r)=\frac{1}{L\sqrt{N}}\Sum\hq e^{iQnd}e^{i\bm q\cdot\bm r},
\end{align}
with $L$ the lateral size of the membrane. Assuming periodic boundary
conditions in all directions, the wavevectors are quantified according
to $Q=2\pi m /(Nd)\in[-\pi/d,\pi/d[$ and $\bm q=(q_x,q_y)= (2\pi m/L,
2\pi \ell/L)$, with $m,\ell \in\mathds{Z}$. The elastic Hamiltonian
takes then the simple form~\cite{safran_book}:
\begin{equation}
\mathcal{H}=\Sum\frac{1}{2}\Aq\hq\hmq\,,
\end{equation}
with
\begin{equation}
\Aq=q^4+2(1-\cos Qd).
\end{equation}

\subsection{Orders of magnitude}

Typical values for the elastic parameters of lipid and surfactant
lamellar phases are given below and  listed in Table~\ref{val}.

\subsubsection{Lipid membranes}\label{omlipids}

For lamellar phases made of egg PC lipids~\cite{Petrache98}, the elastic
constants are $\kappa\simeq\SI{0.5e-19}{J}$, and typically
$B\simeq\SI{1e15}{J.m^{-4}}$ for $w\simeq\SI{1}{nm}$, with
$\delta\simeq\SI{4}{nm}$ and $d\simeq\SI{5}{nm}$. We thus obtain
$\xi=(\kappa/B)^{1/4}\simeq\SI{2.7}{nm}$ and
$\xi_\parallel=\sqrt{\kt}/(B\kappa)^{1/4}\simeq\SI{0.75}{nm}$, yielding
in dimensionless form $\delta\simeq5.3$, $w\simeq1.3$ and $d\simeq6.6$.

\subsubsection{Surfactant membranes}\label{omsurf}

For the $\mathrm{C}_{12}\mathrm{E}_5$/hexanol/water
system~\cite{Freyssingeas96,Beneut08}, with typically a
hexanol/$\mathrm{C}_{12}\mathrm{E}_5$ ratio of $0.35$ and a membrane
fraction of $\phi\simeq7\%$, the elastic constants are
$\kappa\simeq\SI{3.7e-21}{J}$ and $B\simeq\SI{6e8}{J.m^{-4}}$, with
$\delta\simeq\SI{2.9}{nm}$, $d=\delta/\phi\simeq\SI{41.5}{nm}$ and
$w\simeq\SI{38.5}{nm}$. We thus obtain
$\xi=(\kappa/B)^{1/4}\simeq\SI{50}{nm}$,
$\xi_\parallel=\sqrt{\kt}/(B\kappa)^{1/4}\simeq\SI{52}{nm}$, yielding in
dimensionless form $\delta\simeq0.055$, $w\simeq0.74$ and $d\simeq0.79$.

\begin{table}[t]
\small
  \caption{\ Elastic parameters for typical lipid and surfactant
  membranes. The last three lengths are in dimensionless units.}
  \label{tbl:example}
  \begin{tabular*}{0.48\textwidth}{@{\extracolsep{\fill}}lllllll}
    \hline
    & $\kappa$~(\si{J}) & $B~(\si{J.m^{-4}})$ & $\xi;\,\xi_\parallel$~(\si{nm}) & $\delta$ & $w$ & $d$ \\
    \hline
    Egg PC$^{\phantom{1^2}}$ & $\SI{0.5e-19}{}$ & $\SI{1e15}{}$ & 2.7;\,0.75 &5.3 & 1.3 & 6.6\\
    $\mathrm{C}_{12}\mathrm{E}_5$ & $\SI{3.7e-21}{}$ & $\SI{6e8}{}$ & 50;\,52 & 0.06 & 0.74 & 0.79\\
    \hline
  \end{tabular*}
  \label{val}
\end{table}

\subsection{Fluctuations}

The fluctuations of the lamellar phase (in the absence of colloids) are
obtained in a standard way by adding an external field $J_{Q,\bm q}$ to
the partition function~\cite{Chaikin_book}:
\begin{align}
Z[J]&=\int\left(\prod_{n=0}^{N-1}\mathcal{D}[h_n]\right)
\exp\left(-\mathcal{H}+\Sum J_{-Q,-\bm q}\hq\right)
\nonumber\\&
=Z_0\exp\left(
\Sum
\frac{J_{Q,\bm q}J_{-Q,-\bm q}}{\Aq}
\right),
\end{align}
where $Z_0$ is the partition function of the lamellar phase. We shall
denote by $\langle\ldots\rangle$ the statistical average over the
membrane fluctuations. By differentiation, we obtain
\begin{align}\label{corFourier}
\langle \hq h_{Q',\bm q'} \rangle=\left.\frac{\partial^2\ln Z}
{\partial J_{-Q,-\bm q}\partial J_{-Q',-\bm q'}}\right|_{J=0}
=\frac{\delta_{Q+Q'}\delta_{\bm q+\bm q'}}{\Aq}.
\end{align}
The \textit{gap} between layers $p$ and $p+1$, at position $\bm r$, is
given by
\begin{equation}
\label{eq:Hp}
H_p(\bm r)=w+h_{p+1}(\bm r)-h_p(\bm r).
\end{equation}
Its average $\langle H_p(\bm r)\rangle$ is $w$. Using eqns~(\ref{hn})
and (\ref{corFourier}), we obtain its correlation function
\begin{align}
\langle H_0(\bm 0) H_p(\bm r)\rangle-w^2&=
\frac{2}{NL^2}\Sum \frac{1-\cos Qd}{A(Q,q)}e^{iQpd}e^{i\bm q\cdot\bm r}
\nonumber\\
&=\frac{G_p(r)}{2\pi},
\end{align}
where the factor $2\pi$ was introduced for later convenience.
In the thermodynamic limit, 
\begin{equation}
\label{eq:Gp}
G_p(r)=\frac{1}{\pi}\int_{-\pi}^{\pi}\!d\phi\int_0^\infty\!dq\,
\frac{q(1-\cos\phi)\cos(p\phi)\BesselJ_0(qr)}{q^4+2(1-\cos\phi)},
\end{equation}
with, in particular, $G_0(0)=1$, $G_0(r)=2\BesselJ_1(r)\BesselK_1(r)$,
where $\BesselJ_1$ and $\BesselK_1$ are Bessel functions
(see the Appendix), $G_p(0)=1/(1-4p^2)$ and $G_p(\infty)=0$.

It follows that the standard deviation of the gap, or, equivalently of the
layer spacing, is given by $\sigma=\sqrt{G_0(0)/(2\pi)}=1/\sqrt{2\pi}$.
In dimensionful form, this gives $\sigma=\xi\sqrt{\kt/(2\pi\kappa)}$ (in
agreement with ref.~\citenum{Petrache98}). Note that our Gaussian
Hamiltonian~(\ref{eq:H}) takes into account the repulsion of the layers
by means of a soft harmonic repulsive potential. Since
the layers cannot physically interpenetrate, the consistency
of the model requires $w\gtrsim \sigma$, i.e., in dimensionless form,
$w\gtrsim 1/\sqrt{2\pi} \simeq 0.4$. Note that this condition is true
for the parameters given above (see Table~\ref{val}).

Let us also compute the correlation between the membrane gaps and the
layer displacements, that will  show up in the calculation of the
deformation of the lamellar phase induced by the colloids:
\begin{align}
\Gamma_p(r) =
2\pi\langle
H_0(\bm 0)h_p(\bm r)
\rangle=
\frac{2\pi}{NL^2}\Sum \frac{e^{-iQd}-1}{A(Q,q)}e^{iQpd}e^{i\bm q\cdot\bm r}.
\label{eq:Hh}
\end{align}
In the thermodynamic limit, 
\begin{align}
\Gamma_p(r)&=\frac{1}{2\pi}\int_{-\pi}^{\pi}\!d\phi\int_0^\infty\!dq\,
\frac{q(e^{-i\phi}-1)e^{ip\phi}\BesselJ_0(qr)}{q^4+2(1-\cos\phi)},
\end{align}
with, in particular, $\Gamma_p(0)=1/(4p-2)$ and $\Gamma_p(\infty)=0$.

\section{Interactions between colloidal particles}
\label{sec:inter}

Sens and Turner studied the interactions between particles in lamellar
phases in a series of
papers~\cite{Sens97Jphys,Turner97PRE,Turner98PRE,Sens01EPJE}. They
described the particles by pointlike couplings inducing either a local
pinching, a local stiffening or a local curvature of the membrane.
Dealing with both thermotropic and lyotropic smectics, they used the
three-dimensional smectic elasticity expressed in terms of a continuous
layer displacement function~\cite{deGennes_book}. This approach yields
the asymptotic interaction between colloids and thus the collective
phase behavior when the colloids are dispersed, but the obtained
interaction has a peculiar divergence for colloids in the same layer.

An approximate cure to this problem was proposed in
ref.~\citenum{Tovkach13PRE} by the introduction of a high wavevector
cutoff of the order of the inverse smectic spacing. Another difficulty,
inherent to linear couplings, is that the fluctuation corrections to the
interactions are not accounted for. For smectics, since colloids are
always significantly larger than the smectic period, modeling a colloid
in a more realistic manner requires introducing a multipolar
development~\cite{Turner98PRE}, or an effective coat larger than the
particle where the deformation is small enough to use a multipolar
approach~\cite{Tovkach13PRE}. Choosing the multipoles coefficients is
difficult, however, in particular because enforcing strict boundary
conditions make them in general dependent on the distances between the
colloids. Finally, as  shown in ref.~\citenum{Santangelo03PRL},
rotational invariance and the associated non-linear effects can yield
important modifications in the far-field deformations and thus in the
interaction potentials.

In this study, as discussed in Section~\ref{sec:intro}, we use a different
approach, based on a discrete model for the layers. This approach
applies to lyotropic lamellar phases, but not to continuous smectic
phases. It has  the advantage, however, that the colloid-lamellar
coupling is taken into account in an almost exact manner. We therefore
expect to obtain reliable interactions at all separations, including in
particular the fluctuation-induced corrections.

\subsection{Membrane-binding colloids}\label{sec:binding}

Let us start by considering colloids that adhere strongly to the
neighboring bilayers, as the colloid on the right in Fig.~\ref{schema}.
We consider a first colloid of diameter $a$
binding to layers $n=0$ and $n=1$ at the in-plane position $(x,y)=\bm
0$, and a second one of diameter $b$ binding to layers $n=p$ and $n=p+1$
at the in-plane position $(x,y)=\bm R$. We model their binding as a
simple constraint on the gaps between their neighboring membranes on the
axis normal to the undeformed membranes:
\begin{align}
H_0(\bm 0)=a,\quad H_p(\bm R)=b.
\end{align}
The partition function of the system (at fixed projected positions, $\bm
0$ and $\bm R$, of the colloids) is therefore given by
\begin{align}
Z_\mathrm{bind}=\int\left(\prod_{n=0}^{N-1}\mathcal{D}[h_n]\right)
\delta(H_0(\bm 0)-a)\,\delta(H_p(\bm R)-b)\,e^{-\mathcal{H}} .
\end{align}
Using the Fourier representation both of the delta functions and of the
layer displacements yields:
\begin{align}
Z_\mathrm{bind}=\int\!&
\left(\prod_{n=0}^{N-1}\mathcal{D}[h_n]\right)\frac{d\lambda}{2\pi} \frac{d\mu}{2\pi}\,
e^{i\lambda(w-a)+i\mu(w-b)}
\nonumber\\&
\times e^{-
\Sum\left[\frac{1}{2}\Aq\hq\hmq\,-\hq\Smq
\right]},
\label{eq:ZbindJ}
\end{align}
with
\begin{equation}
\Sq=
i\lambda\frac{e^{-iQd}-1}{L\sqrt{N}}
+i\mu\frac{e^{-iQd}-1}{L\sqrt{N}} e^{-iQpd}e^{-i\bm q\cdot\bm R}
+\Jq,
\label{eq:SJ}
\end{equation}
where we have added an external field $\Jq$ that will be used to compute
the average deformation of the lamellar phase in the presence of the
colloids. Performing the Gaussian integrals, and discarding irrelevant
constant factors, yields
\begin{align}
Z_\mathrm{bind}&=\int\!d\lambda\,d\mu\,
e^{i\lambda(w-a)+i\mu(w-b)}
e^{\frac{1}{2}\Sum\frac{1}{\Aq}\Sq\Smq}
\nonumber\\&
=(\det M)^{-1/2}e^{-\frac{1}{2}(s,s')M^{-1}(s,s')^T}
\nonumber\\
&~~~~~~~\times
e^{\frac{1}{2}\Sum\frac{1}{\Aq}\Jq\Jmq},
\end{align}
where $M(R,p)$ is a symmetric $2\times2$ matrix with elements:
\begin{align}
&M_{11}=M_{22}=\frac{2}{NL^2}\Sum\frac{1-\cos Qd}{\Aq}=\frac{G_0(0)}{2\pi},
\nonumber\\&
M_{12}=\frac{2}{NL^2}\Sum\frac{1-\cos Qd}{\Aq}\,\cos(Qpd+\bm q\cdot\bm R)=\frac{G_p(R)}{2\pi},
\end{align}
where we recognize the correlation function of the layer spacing, and
\begin{align}
s&=w-a+\frac{1}{L\sqrt{N}}\Sum\Jq\frac{e^{iQd}-1}{\Aq},
\\
s'&=w-b+\frac{1}{L\sqrt{N}}\Sum\Jq\frac{e^{iQd}-1}{\Aq}
e^{iQpd}e^{i\bm q\cdot\bm R}.
\end{align}

\begin{figure*}[t]
\centerline{\includegraphics[width=1.9\columnwidth]{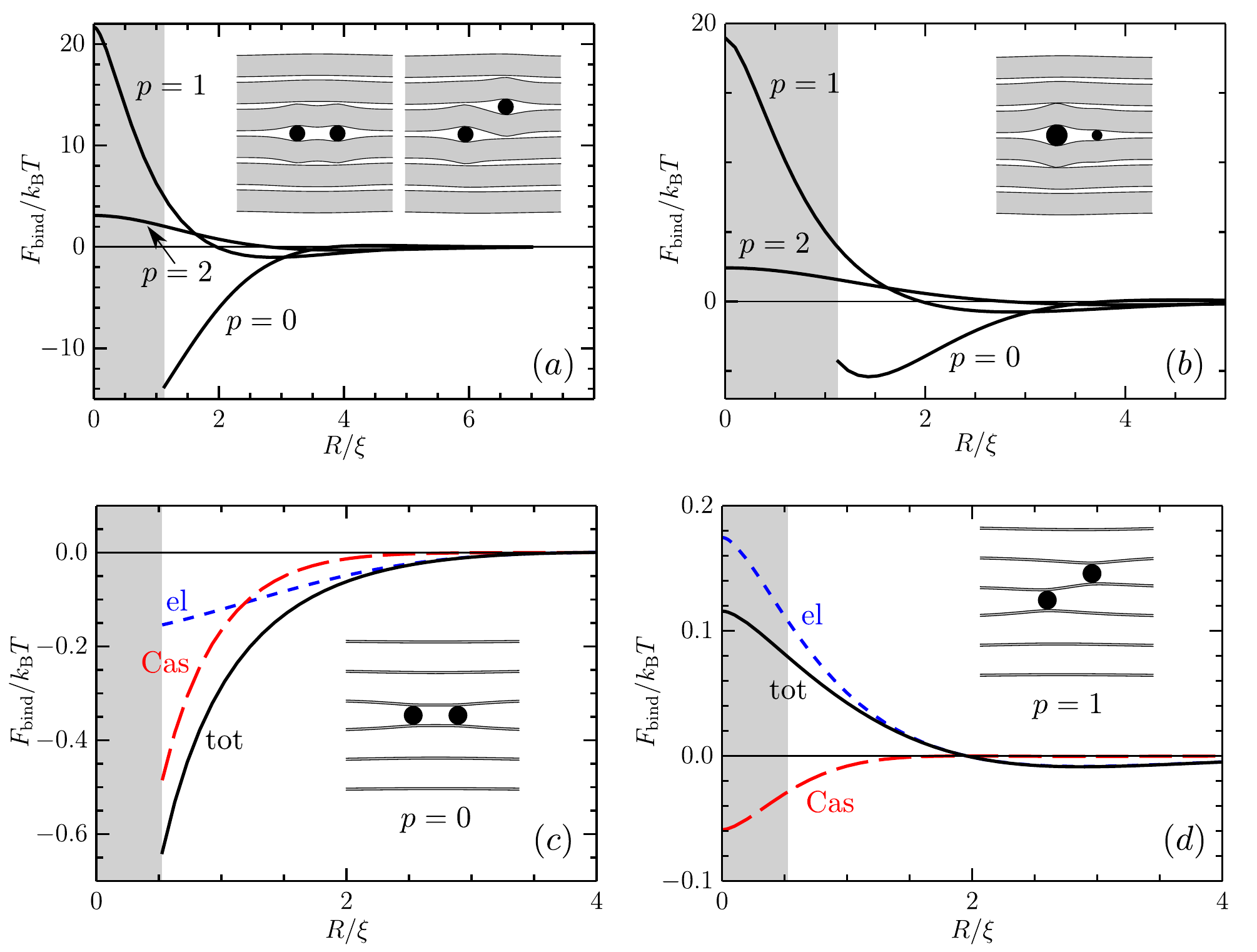}}
\caption{(a) Interaction energy as a function of separation for
membrane-binding colloids with diameters $a=b=3w$ in a lamellar phase
with the parameters of egg PC (see Section~\ref{omlipids}). The Casimir
contribution to the  interaction is less than $1\%$ of the elastic part.
Colloids in the same layer ($p=0$), one layer apart ($p=1$), two layers
apart ($p=2$). Insets: Deformations of the lamellar phase, calculated
numerically using eqn~(\ref{calculatedshape}), for $R/d=1.5$ and
$N=100$. The membranes, in gray, and the colloids, in black, are
represented at scale. (b) Same as (a) for colloids of diameter $a=4w$
and $b=2w$. (c) Interaction energy as a function of separation for
binding colloids of diameter $a=b=0.7w$ placed in the same layer in a
lamellar phase with the parameters of $\mathrm{C}_{12}\mathrm{E}_5$ (see
Section~\ref{omsurf}). Black solid line (tot): total interaction
$F_\mathrm{bind}$. Blue dashed line (el): elastic interaction
$F_\mathrm{bind}^\mathrm{el}$. Red long-dashed line (Cas): Casimir
interaction $F_\mathrm{bind}^\mathrm{Cas}$. Inset: numerically
calculated deformation of the lamellar phase, using
eqn~(\ref{calculatedshape}), for $R/d=1.5$ and $N=100$. The membranes
and the colloids are represented at scale. (d) Same but for colloids one
layer apart ($p=1$).} \label{fig-binding}
\end{figure*}

\subsubsection{Interaction free energy and average deformation}

Taking the thermodynamic limit $N\to\infty$ and $L\to\infty$, we obtain
for $\Jq=0$, apart from an irrelevant constant factor,
\begin{align}
\label{resultZbind}
Z_\mathrm{bind}&=
\left(1-G_p(R)^2\right)^{-1/2}
\nonumber\\
&\times
e^{\displaystyle-\pi
\frac{(a-w)^2+(b-w)^2-2G_p(R)(a-w)(b-w)}{1-G_p(R)^2}}
\end{align}
The total free energy $-\ln(Z_\mathrm{bind})$ of the system yields,
after subtracting the value for infinitely separated colloids, the
interaction free energy of the two colloids:
\begin{align}
\label{eq:Fbind}
&F_\mathrm{bind}(R,p)=
F_\mathrm{bind}^\mathrm{{Cas}}(R,p)+F_\mathrm{bind}^\mathrm{el}(R,p)
\\
&F_\mathrm{bind}^\mathrm{{Cas}}=\frac{1}{2}\ln\left(1-G_p(R)^2\right),
\label{FC}
\\
&F_\mathrm{bind}^\mathrm{el}=-
\frac{2\pi G_p(R)}{1+G_p(R)}
\left[(a-w)(b-w)-\frac{1}{2}\frac{G_p(R)(a-b)^2}{1-G_p(R)}\right].
\label{Fel}
\end{align}

In order to get the dimensionful form of these interactions, one has to
multiply these expressions by $\kt$ and add an extra factor
$\sqrt{\kappa B}/\kt$ in front of $(w-a)(w-b)$ and $(a-b)^2$. The
interaction $F_\mathrm{bind}^\mathrm{{Cas}}$, which is thus directly
proportional to the temperature, is a Casimir-like interaction, caused
by the restriction of the fluctuations induced  by the binding of the
colloids. The interaction $F_\mathrm{bind}^\mathrm{el}$ is an athermal
``elastic'' interaction, proportional to $\sqrt{\kappa B}$ (it depends on
temperature only through $\kappa$ and $B$) and it is caused by the
deformation of the layers induced by the colloids. Note that if $a=b=w$,
in which case the colloids do not deform the layers, the elastic
interaction vanishes while the fluctuation-induced Casimir interaction
remains.

The average deformation of the lamellar phase is given by
\begin{align}
\langle\hmq\rangle_\mathrm{bind}&=\left.\frac{\partial\ln Z_\mathrm{bind}}{\partial\Jq}\right|_{J=0}
\nonumber\\
&=-\frac{1}{2}\left.\frac{\partial\left[(s,s')M^{-1}(s,s')^T\right]}{\partial\Jq}\right|_{J=0},
\end{align}
yielding
\begin{align}
\label{calculatedshape}
\langle h_n(\bm r)\rangle_\mathrm{bind}&=
\frac{a-w-G_p(R)(b-w)}{1-G_p(R)^2}
\Gamma_n(r)
\nonumber\\&
+\frac{b-w-(a-w)G_p(R)}{1-G_p(R)^2}
\Gamma_{n-p}(|\bm r-\bm R|),
\end{align}
where $\Gamma_n$ is the correlation function~(\ref{eq:Hh}).  The
deformation induced by just one colloid, of diameter $a$, is obtained by
taking the first term in the right-hand side of
eqn~(\ref{calculatedshape}) for $G_p(R)=0$ (infinite separation) and
is given by
\begin{align}
\langle h_n^{(1)}(\bm r)\rangle_\mathrm{bind}=
(a-w)\Gamma_n(r).
\end{align}
Note that the deformation set by two colloids is not simply the
superposition of the deformations set by each individual colloid.
This non-linearity comes from the membrane thickness constraint
imposed by the particles.

The deformation above a single colloid, of diameter $a$, placed between
layers $0$ and $1$, is therefore given by \begin{align}
\label{above}
\langle h_n^{(1)}(\bm 0)\rangle_\mathrm{bind}=\frac{a-w}{4n-2}\,.
\end{align}
It is independent of the elastic constants since $a$, $w$, and $h_n$
are normalized with respect to the same length.

For the consistency of our model, we must verify that the gap between
the membranes bound to the colloids and the adjacent ones remain positive despite the
deformation. In particular, we must have $\langle H_1(\bm 0)\rangle>0$.
Given eqns (\ref{above}) and~(\ref{eq:Hp}), this yields the consistency
condition $0<a<4w$ for a colloid of diameter $a$.

\begin{figure*}
\centerline{\includegraphics[width=1.6\columnwidth]{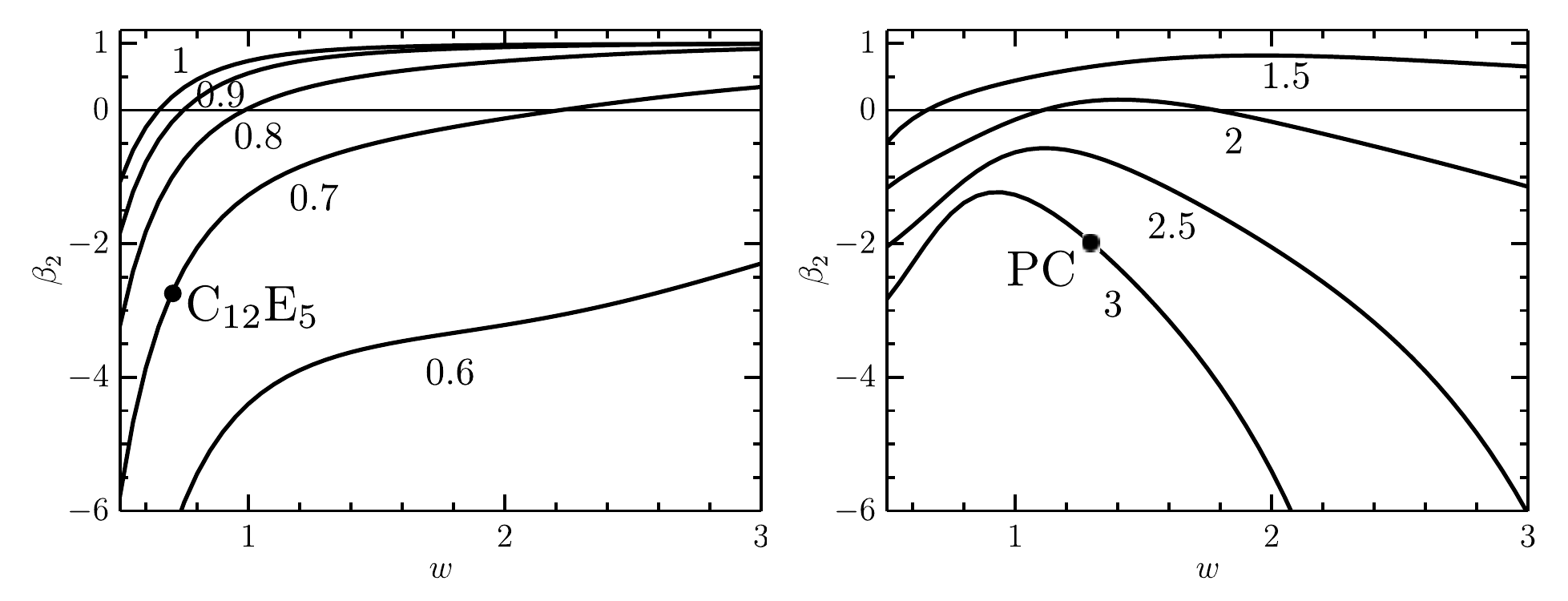}}
\caption{Normalized second virial coefficient $\beta_2$ as a function of
water thickness for the values of the ratio $a/w$ indicated on the
curves. The egg PC and $\mathrm{C}_{12}\mathrm{E}_5$ systems
corresponding to Fig.~\ref{fig-binding} are indicated by the dots.}
\label{B2}
\end{figure*}

\subsubsection{Typical results}

Lamellar phases made with lipids have a layer spacing typically
comparable, or even smaller, than the membrane thickness
$\simeq\SI{4}{nm}$, so that only nano-colloids will fit in such systems
(see Section~\ref{omlipids}).  We show in Fig.~\ref{fig-binding}a the
typical interaction energy between membrane-binding colloids and the
corresponding lamellar phase deformation. The values correspond to egg
PC lipids with colloids of diameters $\simeq\SI{3}{nm}$. Two colloids in
the same layer attract each other, while colloids in different layers
repel one another. The maximum interaction energies are large with
respect to $\kt$. When the colloids are separated by more than one empty
layer, their interaction becomes negligible compared to $\kt$. For such
lipid lamellar phases the Casimir component of the interaction energy is
always negligible with respect to the elastic one. In
Fig.~\ref{fig-binding}b we show the corresponding interaction energy and
associated lamellar phase deformation for two colloids of different
radiuses.

Lamellar phases made of surfactants can have a much larger layer spacing
(see Section~\ref{omlipids}), so that larger colloids can fit in. They also
have weaker elastic constants, so that fluctuation effects are larger.
We show in Figs.~\ref{fig-binding}c and~\ref{fig-binding}d the typical
interaction energy and the corresponding lamellar phase deformation. The
values correspond to $\mathrm{C}_{12}\mathrm{E}_5$ surfactants with
colloids of diameter $\simeq\SI{27}{nm}$, as in the experiments of
ref.\citenum{Beneut08}. The behaviours are similar to those of lipid
membranes, but the energies are much smaller. Also the contribution of
the Casimir interaction is no longer negligible.

\subsubsection{Second virial coefficient}

For membrane-binding colloids of diameter $a$ dispersed in a lamellar
phase, the second virial coefficient is given (in dimensionless form) by
\begin{align}
B_2=\frac{1}{2}\int d^2R\,dz\left(1-e^{-F_\mathrm{bind}(\bm R,p=z/d)}\right),
\end{align}
where the coordinate $z$ is the vertical position of a colloid. Taking
into account the discrete nature of the layers, we make the replacement
$\int f(z) dz \rightarrow d \sum_p f_p$, thus obtaining
\begin{equation}
B_2=\frac{\pi a^2d}{2}\beta_2\,,
\end{equation}
with the normalized second virial coefficient
\begin{equation}
\beta_2=1+\sum_{p=-\infty}^\infty\int_{\delta_{p,0}}^\infty\!dr\,2r\left(
1-\frac{\exp\left(\frac{2\pi(a-w)^2}{1+G_p(ar)^{-1}}\right)}{\sqrt{1-G_p(ar)^2}}
\right).
\end{equation}
In the last equation, we have used eqns~(\ref{FC}) and~(\ref{Fel}) and
we have taken into account the excluded volume interaction between
colloids in the same layer ($p=0$). Note that with our normalization the
pure hard core interaction corresponds to $\beta_2=1$.

The numerically calculated values of $\beta_2$ are shown in
Fig.~\ref{B2} as a function of the water thickness and colloid size. As
we always have $\beta_2<1$, the interaction is always globally
attractive. When $0<\beta_2<1$, however, it doesn't prevail over the
hard core, indicating  a globally stable colloid dispersion.

\begin{figure*}
\centerline{\includegraphics[width=1.9\columnwidth]{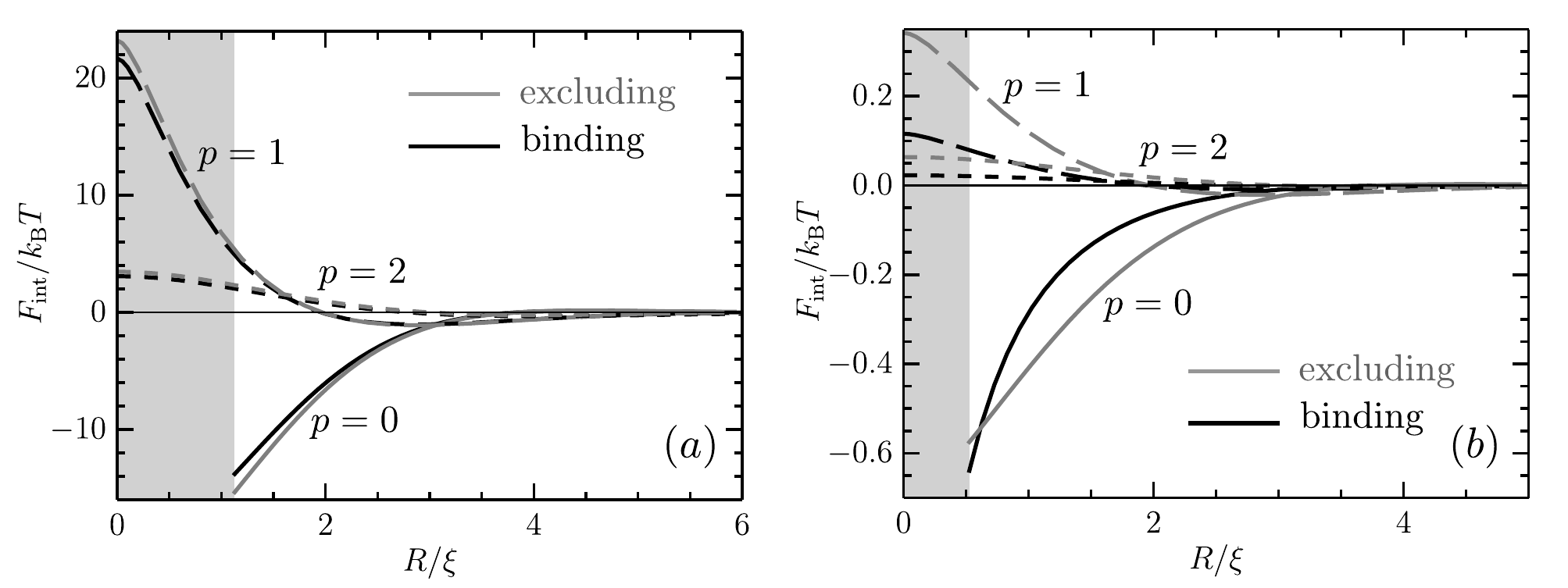}}
\caption{Comparaison between the interaction energies of
membrane-excluding colloids (gray lines) and of membrane-binding
colloids (black lines, same curves as in Fig.~\ref{fig-binding}a, c and
d). Solid lines: colloids in the same layer ($p=0$), long-dashed lines:
colloids one layer appart ($p=1$), dashed-line: colloids two layers
apart ($p=2$). (a) Lamellar phase with the parameters of egg PC and
colloids of diameter $a=b=3w$. (b) Lamellar phase with the parameters of
$\mathrm{C}_{12}\mathrm{E}_5$ and colloids of diameter $a=b=0.7w$. }
\label{fig-bind-ex}
\end{figure*}

\subsection{Membrane-excluding colloids}\label{sec:excluding}

Let us now consider colloids that interact with the membranes only
through excluded volume forces, as the colloid on the left in Fig.~\ref{schema}.
We take a first colloid of diameter $a$
placed between layers $n=0$ and $n=1$ at the in-plane position
$(x,y)=\bm 0$, and a second one of diameter $b$ placed between layers
$n=p$ and $n=p+1$ at the in-plane position $(x,y)=\bm R$.  We model
their presence in between the layers by imposing that the gaps between
their neighboring membranes, on the axis normal to the undeformed
membranes, cannot be smaller than their diameter. Such a constraint
corresponds to the Hamiltonians of the infinite well type:
\begin{align}
&\mathcal{H}_a(z_a)=
\begin{cases}
0&
\text{if~~}
z_a\in[z_0(\bm 0)+\frac{\delta+a}{2},z_1(\bm 0)-\frac{\delta+a}{2}],
\cr
+\infty&\text{otherwise},
\end{cases}
\\
&\mathcal{H}_b(z_b)=
\begin{cases}
0&
\text{if~~}
z_b\in[z_p(\bm R)+\frac{\delta+b}{2},z_{p+1}(\bm R)-\frac{\delta+b}{2}],
\cr
+\infty&\text{otherwise,}
\end{cases}
\end{align}
where $z_a$ (resp.\ $z_b$) is the height of the particle of diameter $a$
(resp.\ $b$) and $z_p(\bm r)=pd+h_p(\bm r)$ is the height of the center
of the membrane number $p$ at the in-plane position $\bm r$.

The partition function, at fixed projected positions $\bm 0$ and $\bm R$
of the colloids, is then given by
\begin{align}
\label{Zex}
Z_\mathrm{ex}=\int
\left(\prod_{n=0}^{N-1}\mathcal{D}[h_n]\right) dz_a\, dz_b\,
e^{-\left[\mathcal{H}+\mathcal{H}_a(z_a)+\mathcal{H}_b(z_b)\right]}\,.
\end{align}
Integrating the Boltzmann weights associated to the infinite wells gives simply
\begin{align}
\int\!dz_a\,e^{-\mathcal{H}_a(z_a)}
&=\left(H_0(\bm 0)-a\right) \Theta\left(H_0(\bm 0)-a\right),
\\
\int\!dz_b\,e^{-\mathcal{H}_b(z_b)}
&=\left(H_p(\bm R)-b\right)\Theta\left(H_p(\bm R)-b\right),
\end{align}
where $H_0(\bm 0)-a$ (resp.\ $H_p(\bm R)-b$) is the gap available to the
first (resp.\ second) colloid and the Heaviside functions $\Theta$ are
such that the integrals vanish when the gaps are smaller than the
colloids diameters. Using the relation $\Theta(x-a)=\int_a^\infty
dg\delta(x-g)$, we can map the problem onto that of binding colloids,
yielding
\begin{align}
Z_\mathrm{ex}(R,p)=\int_a^\infty\!dg\int_b^\infty\!dg'\,(g-a)(g'-b)\,
Z_\mathrm{bind}(g,g'),
\end{align}
where $Z_\mathrm{bind}(g,g')$ is the partition function for binding
colloids of diameters $g$ and $g'$, obtained by replacing $a$ and $b$ by
$g$ and $g'$ in eqn~\ref{resultZbind}. This expression can be understood
if one thinks of integrating first over the gaps $g\in[a,\infty]$ and
$g'\in[b,\infty]$ of the layers surrounding the colloids, then over the
other degrees of freedom at fixed gaps: the integration of the membrane
degrees of freedom at fixed gaps gives the partition
function~(\ref{resultZbind}) for membrane-binding colloids, while the
integration over the particle positions gives the entropic contributions
$g-a$ and $g'-b$.

\subsubsection{Interaction free energy and average deformation}

The interaction free energy for two membrane-excluding colloids is
therefore given by
\begin{equation}
\label{eq:Fexcl}
F_\mathrm{ex}(R,p)=-\ln \frac{Z_\mathrm{ex}(R,p)}{Z_\mathrm{ex}(+\infty,p)},
\end{equation}
which can be easily calculated numerically by a double integration. Note
that it is no longer possible here to extract separately an elastic
contribution and a Casimir one.

The average deformation of the layers can be calculated by adding to the
partition function~(\ref{Zex}) an external field $J$ as in eqns
(\ref{eq:ZbindJ})--(\ref{eq:SJ}). From the relation
$\langle\hmq\rangle_\mathrm{ex}=(1/Z_\mathrm{ex})\partial
Z_\mathrm{ex}/\partial\Jq|_{J=0}$, using $\partial
Z_\mathrm{bind}(g,g')/\partial\Jq|_{J=0}=\langle\hmq\rangle_\mathrm{bind}\times
Z_\mathrm{bind}(g,g')$ yields in direct space:
\begin{align}
\label{defex}
\langle h_n(\bm r)\rangle_\mathrm{ex}&=
\frac{1}{Z_\mathrm{ex}} \int_a^\infty\!dg\int_b^\infty\!dg'\,
(g-a)(g'-b)
\nonumber\\&\times
\langle h_n(\bm r)\rangle_\mathrm{bind}\,
Z_\mathrm{bind}(g,g'), \end{align} with $\langle h_n(\bm
r)\rangle_\mathrm{bind}$ the average layer deformation at projected
position $\bm r$ for two colloids of diameters $g$ and $g'$.  This
expression can also be understood intuitively, since $\langle h_n(\bm
r)\rangle_\mathrm{bind}\, Z_\mathrm{bind}$ is the integral over all the
microstates corresponding to fixed gaps $g$ and $g'$ of $h_n(\bm r)$
multiplied by $\exp(-\mathcal{H})$.

Since for one isolated binding colloid of radius $g$ we have $\langle
H_0\rangle_\mathrm{bind}=g$ and $Z_\mathrm{bind}=C\exp[-\pi(g-w)^2]$
(see eqn~(\ref{resultZbind})), the average gap $\langle H_0^{(1)}\rangle$ set
by one hard-core colloid of radius $a$ is given by
\begin{align}
\langle
H_0^{(1)}\rangle_\mathrm{ex} &=\frac{ \int_a^\infty
(g-a)g\,e^{-\pi(g-w)^2} }{ \int_a^\infty (g-a)\,e^{-\pi(g-w)^2} }
\nonumber\\ &=w+\left( 2\pi(w-a)
+\frac{2e^{-\pi(a-w)^2}}{\mathrm{erfc}[\sqrt{\pi}(a-w)]} \right)^{-1}.
\end{align}

\begin{figure*}[t]
\centerline{\includegraphics[width=1.9\columnwidth]{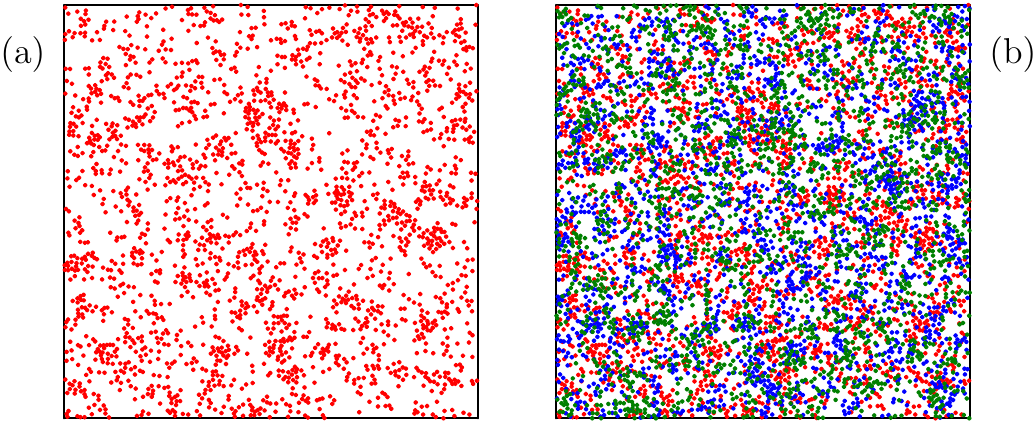}}
\caption{Typical snapshot of a Monte Carlo simulation of the membrane-excluding colloids (after
equilibration) for the parameters of $\mathrm{C}_{12}\mathrm{E}_5$ (see
Section~\ref{omsurf}) for particles of diameter $a=\SI{27}{\nano\meter}$
and hard-core distance between colloids of $\SI{34}{\nano\meter}$ . The colloid volume fraction is~2\%. 
The colloids in one layer are represented as red disks.
The blue and green disks represent the colloids in the two adjacent layers.  The diameter of the disks corresponds
to the hard-core distance. (a) Colloids in one layer. (b) Red disks: same layer of colloids as in (a); blue and green disks: colloids
in the adjacent two layers for the same snapshot.}
\label{fig:MCv}
\end{figure*}

\subsubsection{Typical results}

In Fig.~\ref{fig-bind-ex} we compare the interaction energies between
membrane-excluding and membrane-binding colloids for egg PC lipids with
colloids of diameters $\simeq\SI{3}{nm}$ and for
$\mathrm{C}_{12}\mathrm{E}_5$ surfactants with colloids of diameter
$\simeq\SI{27}{nm}$, as in Fig.~\ref{fig-binding}.

In the case of egg PC lipids, the colloid diameters are much larger than
the average water thickness. Then, the configurations that are
effectively sampled by the fluctuations do not significantly depend on
whether the colloids stick to the layers or not, as spreading the layers
further away from the colloids costs a large energy. This is why, the
interaction energies for the membrane-excluding and membrane-binding
cases are very close (see Fig.~\ref{fig-bind-ex}a).

Conversely, in the case of $\mathrm{C}_{12}\mathrm{E}_5$ surfactants,
the colloid diameters are slightly smaller than the average water
thickness. Therefore, the interaction between membrane-excluding
colloids is of pure fluctuation (Casimir) origin: in the absence of
fluctuations, the colloids sit anywhere in between the layers without
producing any deformation, whatever their distance. As seen in
Fig.~\ref{fig-bind-ex}b, in this case, the interaction energies for the
membrane-excluding and membrane-binding cases differ significantly, even
though the overall behavior is similar. Due to the various contributions
to the free energies (elastic deformations, entropy associated to the
fluctuations of the membranes and of the colloids) and their differences
in the two situations, it is difficult to get a qualititative
understanding of the interaction energy variations between the two
situations.

\section{Monte Carlo simulations}
\label{sec:MC}

To investigate the equilibrium statistical behavior of the interacting
colloids, we use a Monte Carlo simulation with a Metropolis algorithm.
We simulate only the behavior of the colloidal particles, subjected
to the membrane-mediated interaction that we have
previously computed. Precisely, we model the colloid-lamellar phase system as a finite number $M$ of
stacks of identical colloids orthogonal to $z$-direction, each one
consisting of the same finite number $N$ of particles confined in a disk
of radius $R_d$. We suppose that the particles cannot change stack. To
simplify, we consider only pairwise interactions (i.e., we neglect
multibody effects) and we take into account only the contributions
coming from particles in the same layer and one layer apart. Indeed, as
we saw in Section~\ref{sec:inter}, the interaction decreases rapidly with
the layer separation. Moreover, since the interactions are short-ranged, we do not impose periodic boundary conditions within each layer, but we do use periodic boundary conditions in the $z$-direction, such that a small number ($\simeq 7$) of layers is enough for simulating an infinite system.

We start the simulation by placing the same number $N$ of particles in
each one of the $M$ layers according to a random uniform distribution
respecting a given hard-core minimum distance $a_0$. We then pick at
random one particle and we move it randomly inside a circle of radius
$\epsilon$. We compute the associated variation of the interaction
energy and we accept the movement according to the Metropolis rule,
taking into account the hard-core constraint. The radius $\epsilon$ is
adjusted in order to have an acceptance ratio of $\simeq 50\%$.  The
interaction energy between two colloids is computed according to
eqn~(\ref{eq:Fbind}) [resp. (\ref{eq:Fexcl})] for the membrane-binding
(resp. membrane-excluding) case, where the correlation functions
$G_0(R)$ and $G_1(R)$ [see eqn~(\ref{eq:Gp})] can be expressed
analytically in terms of modified Bessel functions, as shown in the
Appendix.

After equilibration, we characterize the statistical order of the
colloids by means of the structure
factor~\cite{Yang:1999,Constantin:2010}:
\begin{align}
S(\bm q, Q) &= \frac1{MN} \left\langle \left|
\sum_{j,p}\exp\left[i\left(\bm q\cdot\bm R_{jp}
+p Q d\right)\right]
\right|^2\right\rangle,\nonumber\\
&=S_0(\bm q) + 2 \sum_{n=1}^\infty S_n(\bm q) \cos(n Q d)\,,
\label{eq:Sq}
\end{align}
where $\bm R_{jp}$ is the position of the $j$-th particle of the $p$-th
layer and $\bm q$ (resp. $Q$) is the component of the wavevector
parallel (resp.\ perpendicular) to the lipid layers. The structure
factor is proportional to the Fourier transform of the two-particle
correlation function. Note that in eqn~(\ref{eq:Sq}) we neglect the
fluctuations of the colloids in the $z$-direction. The partial structure
factors $S_n(\bm q)$ describe the correlations between particles $n$
layers apart.

For a liquid-like order, the structure factors do not depend on the
orientation of the $\bm q$ vector and thus coincide with their average
with respect to the orientation of $\bm q$:
\begin{equation}
\label{eq:Sn}
S_n(q) = \frac{1}{M N}\left\langle \sum_{i,j=0}^{N-1}\sum_{p=0}^{M-1}
\BesselJ_0\left(q\left|\bm{R}_{ip}-\bm{R}_{j(p+n)}\right| \right)\right\rangle
\,,
\end{equation}
where $q$ is the modulus of $\bm q$ and because of the periodic boundary
conditions in the direction perpendicular to the layers, layers
$p$ and $p+M$ coincide.

\begin{figure}[t]
\centerline{\includegraphics[width=0.9\columnwidth]{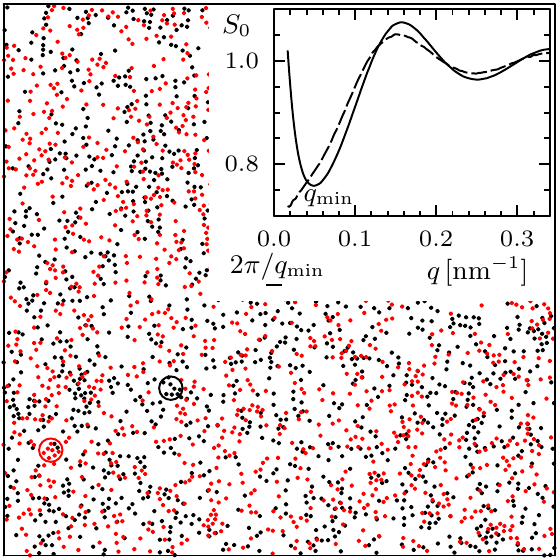}}
\caption{Typical snapshot of a Monte Carlo simulation of the membrane-binding colloids (after
equilibration). The black disks are in one layer and the red disks in an adjacent one.
The diameter of the disks corresponds to the hard core of the colloids.
Inset: structure factor $S_0(q)$ 
inside the layers in the presence of the interparticle interaction (continuous line)
and only with hard core repulsion (dashed line). The position
$q_\mathrm{min}$ of the
minimum of $S_0(q)$ gives an upper estimate of the width of the peak at $q=0$.
The corresponding length $2\pi/q_\mathrm{min}$
is materialized by the bar and by the diameter
of the black and red circles surrounding small clusters in two
adjacent layers. The parameters of
the Monte Carlo simulation correspond to the values given in
Section~\ref{sec:saxs} for a colloid volume fraction of~2\%.}
\label{fig:MC}
\end{figure}

\begin{figure}[t]
\centerline{\includegraphics[width=0.9\columnwidth]{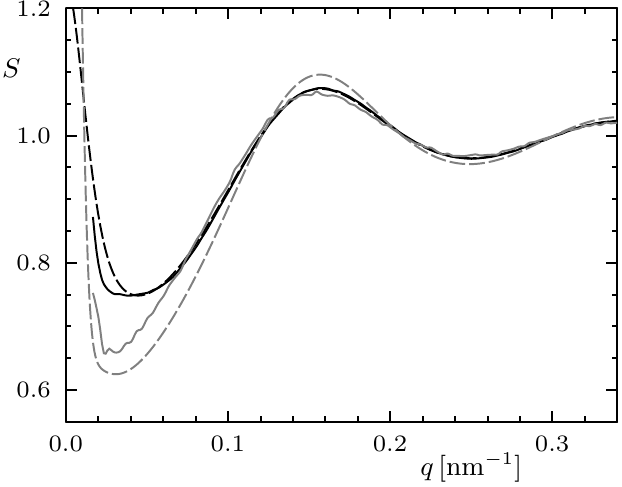}}
\caption{Comparison of the equatorial structure factor $S(q)=S_0(q)+2S_1(q)$
computed with a Monte Carlo simulation (full lines) and the
Percus-Yevick approximation (dashed lines). The black (resp.\ gray)
curves correspond to the membrane-binding (resp.\ membrane-excluding)
case.  The parameters of the Monte Carlo simulation correspond to the
values given in Section~\ref{sec:saxs} for a colloid volume fraction $\phi=2\%$.}
\label{fig:Lado}
\end{figure}

In Fig.~\ref{fig:MCv} we show a typical Monte Carlo
snapshot of three successive layers (blue, red, and green disks)
for membrane-excluding colloids of diameter $\SI{27}{\nano\meter}$
embedded in a lamellar phase with the parameters of $\mathrm{C}_{12}\mathrm{E}_5$.
The corresponding interaction energy is displayed
in Figs.\ \ref{fig-binding}c and \ref{fig-binding}d.
To the membrane-mediated energy we added a hard core
interaction with an effective core diameter of $\SI{34}{\nano\meter}$, as measured in aqueous solution. (see Section\ \ref{sec:saxs}). Clearly, the colloids in
each layer tend to aggregate in large clusters. Moreover, the clusters are
statistically anticorrelated between
adjacent layers: clusters in a layer tends to face voids in the adjacent layers.
This organization originates from the attractive (resp.\ repulsive) character of the
interaction between two colloids sitting in the same layer (resp.\ one
layer apart), as shown in Figs.\ \ref{fig-binding}c and
\ref{fig-binding}d.

\begin{figure*}[t]
	\centerline{\includegraphics[width=1.9\columnwidth]{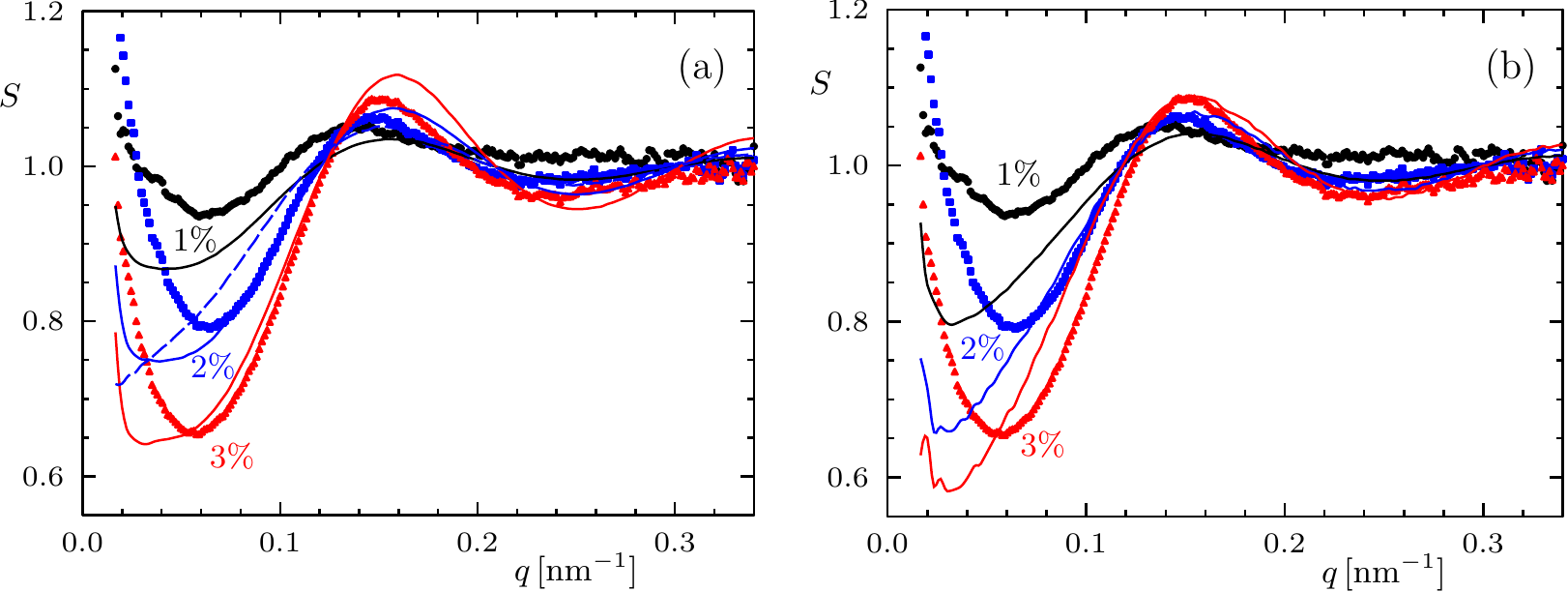}}
	\caption{Experimental in-plane structure factors for silica
		nanoparticles confined in lamellar phases at three different volume
		concentrations $\phi$ \cite{Beneut08} (black circles: $\phi=1\%$; blue squares: $\phi=2\%$; red triangles: $\phi=3\%$) and Monte
		Carlo predictions according to our model (solid lines). (a) Membrane-binding inclusions.
		(b) Membrane-excluding inclusions. Black solid lines: $\phi=1\%$; blue solid lines: $\phi=2\%$; red solid lines: $\phi=3\%$.
		The dashed blue line in (a) is the simulated structure factor for $\phi=2\%$ with only hard-core interactions.
		To obtain convergence, in the membrane-binding [resp.\ membrane-excluding] case, the Monte Carlo averages are performed on
        $10^7$ (resp.\ $3\times 10^8$) steps after equilibration on a system
       consisting of 7 layers having a reduced radius $R=20$ (resp.\ $R=160$). }
	\label{fits}
\end{figure*}

For the same parameters, the membrane-binding colloids tend also to form
clusters, although they are less marked (see Fig.~\ref{fig:MC}).
To assess them, we show  in the inset of Fig.~\ref{fig:MC}
the intra-layer structure factor $S_0$ (solid line). For comparison,
we also show (dashed line) the structure
factor $S_0$ obtained by switching off the interactions, thus taking
into account only the hard core contribution. The first maximum
at $q\simeq \SI{0.16}{{\nano\meter}^{-1}}$ corresponds to the
hard core diameter of the particles. At smaller wavevectors,
the structure factor in the presence of interaction shows a rise
for $q\to 0$ that is absent in the case of hard core only interaction
(dashed line). This can be understood as due to the form factor
of random fluctuating clusters with a distribution of sizes down to
$2\pi/q_\mathrm{min}$, where $q_\mathrm{min}$ is the position
of the minimum of $S_0$ close to $q=0$. Indeed, $q_\mathrm{min}$
gives an upper estimate of the size of the peak at $q=0$.
In the snapshot we have indicated
this size by sourrounding small clusters in two adjacent layers.
This is compatible with the fact that, as
shown in Fig.~\ref{fig-bind-ex}b, the interaction energy for
membrane-binding colloids has the same overall shape, but lower amplitude
in comparison with membrane-excluding ones. Increasing the particle concentration
results in a similar cluster structure in a denser system.

In the literature, the structure factors of pairwise interacting
particles are often calculated in the framework of the Ornstein-Zernicke
relation with the approximate Percus-Yevick closure, using the numerical
method introduced by Lado~\cite{Lado:1967, Lado:1968}. Mapping our
multilayer problem to a multicomponent fluid, as done in
ref.~\citenum{Constantin:2010}, we have computed the equatorial structure factor $S(q)
= S_0(q)+2 S_1(q)$ that can be measured by small-angle X-ray scattering
with a scattering vector contained in the plane of the layers (see Section \ref{sec:saxs}). At small concentrations, we find that
the Percus-Yevick approximation is reasonably good (see
Fig.~\ref{fig:Lado}). However, at higher concentrations for our system
the Lado algorithm does not converge in general.

\section{SAXS}
\label{sec:saxs}

The small-angle X-ray scattering (SAXS) data was acquired as described
in ref.~\citenum{Beneut08}. Briefly, we performed scattering experiments
on oriented lamellar stacks of $\mathrm{C}_{12}\mathrm{E}_5$ surfactants,
with silica particles of nominal diameter of $\SI{27}{\nano\meter}$,
at normal incidence (the beam is parallel to
the smectic director). We recorded the two-dimensional intensity
$I(q=|\mathbf{q}|)$, with $\mathbf{q} = (q_x, q_y)$ the scattering
vector in the plane of the layers and $Q=0$ the scattering vector othogonal to the
layers. The structure factors of the various
samples (see Fig.~\ref{fits}) were obtained as $S(q) = S(q,Q=0) =
I(q)/|F(q)|^2$, where $|F(q)|^2$ is the form factor of the particles,
measured in aqueous suspension.
Note that eqn~(\ref{eq:Sq}) is valid only for the perfect case where the particles do not fluctuate in
the $z$-direction. The thermal fluctuations and the frozen-in defects lead to
a smearing of the diffraction pattern an high $Q$ values, that we describe phenomenologically
by a Lorentzian factor~\cite{Constantin:2010}
\begin{equation}
 S(\bm{q},Q) = S_0(\bm{q},Q) + 2 \frac{\cos(Q d)}{1+\left(Q \sigma\right)^2} S_1(\bm{q})\,,
\end{equation}
where the disorder parameter $\sigma$ has units of length.

\subsection{Membrane-binding predictions}
We compare the measured structure factors with our Monte Carlo
simulations, described in Section~\ref{sec:MC}. For the latter, we adjusted the
radius $R_d$ of the simulation box and the number of layers in order
to have convergence for the correlation functions.
Clearly, using only the hard-core repulsion
(dashed blue line in Fig.~\ref{fits}a), with an effective core diameter of $\SI{34}{\nano\meter}$ describing
the interaction measured in aqueous solution (the core is larger than the nominal diameter due to the electrostatic repulsion),
does not yield a good description of the experimental data on the left of
the structure peak ($q<\SI{0.15}{nm^{-1}}$).

Including the membrane-binding colloid interaction,
with the elastic parameters of $\mathrm{C}_{12}\mathrm{E}_5$ surfactants given in
Table~\ref{tbl:example}, captures quantitatively, with no
adjustable parameters, the experimental points down to the
small-angle increase ($q>\SI{0.05}{nm^{-1}}$). The latter
($q<\SI{0.05}{nm^{-1}}$) is described qualitatively by the complete model,
while it is obviously absent in a hard-core system.

\subsection{Membrane-excluding predictions}
The membrane-excluding model (Fig.~\ref{fits}b) also predicts a small-angle increase,
which is however less important than for the binding case; overall, this model agrees less well
with the experimental data.

\section{Conclusion}
\label{sec:conc}

We treated in detail the interaction between hard spherical inclusions
in lyotropic smectics, for the limiting cases of membrane-excluding and
membrane-binding particles. In both cases, the interaction range is of
the order of the elastic correlation length $\xi=(\kappa/B)^{1/4} =
\sqrt{\lambda d}$ defined in Section~\ref{sec:lamellar}.
For membrane-binding colloids of identical diameters~$a$, the interaction
energy~(\ref{eq:Fbind}) at contact, in the limit $a\ll \xi$,
is approximatively given, in dimensionful form, by
\begin{equation}
 F_\mathrm{bind}^\mathrm{contact}\simeq \frac{\kt}{2} \log C
 +\frac{\pi}{4}\left(C-4\right)\sqrt{B\kappa}\left(a-w\right)^2\,,
\end{equation}
where
\begin{equation}
 C=\left(\frac{a}{\xi}\right)^2\left[\ln\left(\frac{\xi}{a}\right)+\frac{3}{4}+\log 2-\gamma\right]\,,
\end{equation}
with $\gamma\simeq0.577$ the Euler constant. This contact energy varies from tens of $\kt$ for
lipid systems to fractions of $\kt$ in dilute phases of single-chain
surfactants.

For systems of the latter type we compared our predictions to experimental structure factors measured at three concentrations of silica
nanoparticles in a dilute lamellar phase of nonionic surfactant. We obtain semi-quantitative agreement with no adjustable parameters.
Remarkably, this agreement is significantly better for the membrane-binding model than for the membrane-excluding one,
consistent with strong adsorption of these surfactants onto silica surfaces, a result widely accepted in the literature
(see, e.g., the discussion in ref.~\citenum{Sharma:2010}.)

The presence of the particles acts as a constraint on the
membrane fluctuations, leading to an attractive ``Casimir-like''
component of the interaction, which is quite significant (or even
dominant) for the surfactant systems discussed above. 

For strongly attractive systems (with a negative second virial
coefficient), the liquid of particles is unstable with respect to
aggregation. The peculiar nature of the interaction (overall attractive
in the plane of the layers and repulsive across the layers) leads to the
formation of flat and size-limited aggregates. As a concrete
application, one could consider dispersing the particles into a host
lamellar phase with suitably chosen parameters so that they remain well separated, and
then inducing their aggregation by an external stimulus (temperature
change, controlled drying, etc) that increases the interparticle
attraction. The resulting assemblies could then be stabilized by various
strategies \cite{Boles:2016}.

\section*{Acknowledgements}
The SAXS experiments were performed on beamline ID02 at the European
Synchrotron Radiation Facility (ESRF), Grenoble, France. We are grateful
to Pierre Panine at the ESRF for providing assistance in using beamline
ID02. This work was supported by the ANR under contract MEMINT
(2012-BS04-0023).

\section*{Appendix}
\label{sec:appendix}

Performing the integral in $\phi$ in eqn.~\ref{eq:Gp} by means of the
residue theorem, the correlation function inside one layer can be
written as

\begin{equation}
G_0(r) = \int_0^\infty \left(q-\frac{q^3}{\sqrt{q^4+4}}
\right)\BesselJ_0(qr)\,dq\,.
\end{equation}

We use the integral~\cite{Watson_book}
\begin{equation}
\label{eq:int1}
\int_0^\infty q^{1-\nu}\BesselJ_\nu(q r)\, dq = \frac{r^{\nu-2}}{2^{\nu-1}
\Gamma(\nu)}
\end{equation}
for $\mathop{\textrm{Re}}(\nu)>1/2$, where $\Gamma(\nu)$ is the
gamma function, and, for $\mathop{\textrm{Re}}(\nu)>1/6$,
the integral~\cite{Watson_book}
\begin{equation}
\label{eq:int3}
\int_0^\infty \frac{q^{\nu+3}}{\left(q^4+4\right)^{\nu+1/2}}
\BesselJ_\nu(q r)\, dq = \frac{r^\nu\sqrt{\pi}}{2^{3\nu-1}
\Gamma(\nu+\frac{1}{2})}\BesselJ_{\nu-1}(r)\BesselK_{\nu-1}(r)\,.
\end{equation}
Then, taking the difference between eqn (\ref{eq:int1}) and~(\ref{eq:int3}),
by analytical continuation we get, in the limit $\nu\to 0$,
\begin{equation}
G_0(r) = 2\BesselJ_1(r)\BesselK_1(r) \,.
\end{equation}

Similarly, the correlation function~(\ref{eq:Gp}) one layer apart
can be expressed as
\begin{equation}
G_1(r) = \int_0^\infty \left(\frac{q^5}{2}-\frac{q^3}{\sqrt{q^4+4}}
-\frac{q^7}{2\sqrt{q^4+4}}\right)\BesselJ_0(qr)\,dq\,.
\end{equation}
Proceeding as before, it is then easily found that
\begin{align}
G_1(r) &= \left[\left(\frac{32}{r^3}-\frac{4}{r}\right)\BesselJ_1(r)
-\frac{16}{r^2}\BesselJ_0(r) \right] \BesselK_0(r) \nonumber\\
&+\left[\left(\frac{64}{r^4}-2\right)\BesselJ_1(r)
-\left(\frac{32}{r^3}+\frac{4}{r}\right)\BesselJ_0(r)\right]\BesselK_1(r) \,.
\end{align}

\providecommand*{\mcitethebibliography}{\thebibliography}
\csname @ifundefined\endcsname{endmcitethebibliography}
{\let\endmcitethebibliography\endthebibliography}{}


\begin{mcitethebibliography}{40}
\providecommand*{\natexlab}[1]{#1}
\providecommand*{\mciteSetBstSublistMode}[1]{}
\providecommand*{\mciteSetBstMaxWidthForm}[2]{}
\providecommand*{\mciteBstWouldAddEndPuncttrue}
  {\def\EndOfBibitem{\unskip.}}
\providecommand*{\mciteBstWouldAddEndPunctfalse}
  {\let\EndOfBibitem\relax}
\providecommand*{\mciteSetBstMidEndSepPunct}[3]{}
\providecommand*{\mciteSetBstSublistLabelBeginEnd}[3]{}
\providecommand*{\EndOfBibitem}{}
\mciteSetBstSublistMode{f}
\mciteSetBstMaxWidthForm{subitem}
{(\emph{\alph{mcitesubitemcount}})}
\mciteSetBstSublistLabelBeginEnd{\mcitemaxwidthsubitemform\space}
{\relax}{\relax}

\bibitem[Bisoyi and Kumar(2011)]{Bisoyi:2011}
H.~K. Bisoyi and S.~Kumar, \emph{Chem. Soc. Rev.}, 2011, \textbf{40},
  306--319\relax
\mciteBstWouldAddEndPuncttrue
\mciteSetBstMidEndSepPunct{\mcitedefaultmidpunct}
{\mcitedefaultendpunct}{\mcitedefaultseppunct}\relax
\EndOfBibitem
\bibitem[Pratibha \emph{et~al.}(2009)Pratibha, Park, Smalyukh, and
  Park]{Pratibha:2009}
R.~Pratibha, K.~Park, I.~I. Smalyukh and W.~Park, \emph{Optics express}, 2009,
  \textbf{17}, 19459--19469\relax
\mciteBstWouldAddEndPuncttrue
\mciteSetBstMidEndSepPunct{\mcitedefaultmidpunct}
{\mcitedefaultendpunct}{\mcitedefaultseppunct}\relax
\EndOfBibitem
\bibitem[Liu \emph{et~al.}(2014)Liu, Yuan, and Smalyukh]{Liu:2014}
Q.~Liu, Y.~Yuan and I.~I. Smalyukh, \emph{Nano Letters}, 2014, \textbf{14},
  4071--4077\relax
\mciteBstWouldAddEndPuncttrue
\mciteSetBstMidEndSepPunct{\mcitedefaultmidpunct}
{\mcitedefaultendpunct}{\mcitedefaultseppunct}\relax
\EndOfBibitem
\bibitem[Wojcik \emph{et~al.}(2009)Wojcik, Lewandowski, Matraszek, Mieczkowski,
  Borysiuk, Pociecha, and Gorecka]{Wojcik:2009}
M.~Wojcik, W.~Lewandowski, J.~Matraszek, J.~Mieczkowski, J.~Borysiuk,
  D.~Pociecha and E.~Gorecka, \emph{Angewandte Chemie (International Ed. in
  English)}, 2009, \textbf{48}, 5167--5169\relax
\mciteBstWouldAddEndPuncttrue
\mciteSetBstMidEndSepPunct{\mcitedefaultmidpunct}
{\mcitedefaultendpunct}{\mcitedefaultseppunct}\relax
\EndOfBibitem
\bibitem[Coursault \emph{et~al.}(2012)Coursault, Grand, Zappone, Ayeb,
  L{\'e}vi, F{\'e}lidj, and Lacaze]{Coursault:2012}
D.~Coursault, J.~Grand, B.~Zappone, H.~Ayeb, G.~L{\'e}vi, N.~F{\'e}lidj and
  E.~Lacaze, \emph{Advanced Materials}, 2012, \textbf{24}, 1461--1465\relax
\mciteBstWouldAddEndPuncttrue
\mciteSetBstMidEndSepPunct{\mcitedefaultmidpunct}
{\mcitedefaultendpunct}{\mcitedefaultseppunct}\relax
\EndOfBibitem
\bibitem[Lewandowski \emph{et~al.}(2013)Lewandowski, Constantin, Walicka,
  Pociecha, Mieczkowski, and G{\'o}recka]{Lewandowski:2013}
W.~Lewandowski, D.~Constantin, K.~Walicka, D.~Pociecha, J.~Mieczkowski and
  E.~G{\'o}recka, \emph{Chemical Communications}, 2013, \textbf{49},
  7845--7847\relax
\mciteBstWouldAddEndPuncttrue
\mciteSetBstMidEndSepPunct{\mcitedefaultmidpunct}
{\mcitedefaultendpunct}{\mcitedefaultseppunct}\relax
\EndOfBibitem
\bibitem[Liu \emph{et~al.}(2010)Liu, Cui, Gardner, Li, He, and
  Smalyukh]{Liu:2010}
Q.~Liu, Y.~Cui, D.~Gardner, X.~Li, S.~He and I.~I. Smalyukh, \emph{Nano
  Letters}, 2010, \textbf{10}, 1347--1353\relax
\mciteBstWouldAddEndPuncttrue
\mciteSetBstMidEndSepPunct{\mcitedefaultmidpunct}
{\mcitedefaultendpunct}{\mcitedefaultseppunct}\relax
\EndOfBibitem
\bibitem[Venugopal \emph{et~al.}(2011)Venugopal, Bhat, Vallooran, and
  Mezzenga]{Venugopal:2011}
E.~Venugopal, S.~K. Bhat, J.~J. Vallooran and R.~Mezzenga, \emph{Langmuir},
  2011, \textbf{27}, 9792--9800\relax
\mciteBstWouldAddEndPuncttrue
\mciteSetBstMidEndSepPunct{\mcitedefaultmidpunct}
{\mcitedefaultendpunct}{\mcitedefaultseppunct}\relax
\EndOfBibitem
\bibitem[Wang \emph{et~al.}(1999)Wang, Efrima, and Regev]{Wang:1999}
W.~Wang, S.~Efrima and O.~Regev, \emph{The Journal of Physical Chemistry B},
  1999, \textbf{103}, 5613--5621\relax
\mciteBstWouldAddEndPuncttrue
\mciteSetBstMidEndSepPunct{\mcitedefaultmidpunct}
{\mcitedefaultendpunct}{\mcitedefaultseppunct}\relax
\EndOfBibitem
\bibitem[Firestone \emph{et~al.}(2001)Firestone, Williams, Seifert, and
  Csencsits]{Firestone:2001}
M.~A. Firestone, D.~E. Williams, S.~Seifert and R.~Csencsits, \emph{Nano
  Letters}, 2001, \textbf{1}, 129--135\relax
\mciteBstWouldAddEndPuncttrue
\mciteSetBstMidEndSepPunct{\mcitedefaultmidpunct}
{\mcitedefaultendpunct}{\mcitedefaultseppunct}\relax
\EndOfBibitem
\bibitem[Constantin and Davidson(2014)]{Constantin:2014}
D.~Constantin and P.~Davidson, \emph{ChemPhysChem}, 2014, \textbf{15},
  1270--1282\relax
\mciteBstWouldAddEndPuncttrue
\mciteSetBstMidEndSepPunct{\mcitedefaultmidpunct}
{\mcitedefaultendpunct}{\mcitedefaultseppunct}\relax
\EndOfBibitem
\bibitem[Henry \emph{et~al.}(2011)Henry, Dif, Schmutz, Legoff, Amblard,
  Marchi-Artzner, and Artzner]{Henry:2011}
E.~Henry, A.~Dif, M.~Schmutz, L.~Legoff, F.~Amblard, V.~Marchi-Artzner and
  F.~Artzner, \emph{Nano Letters}, 2011, \textbf{11}, 5443--5448\relax
\mciteBstWouldAddEndPuncttrue
\mciteSetBstMidEndSepPunct{\mcitedefaultmidpunct}
{\mcitedefaultendpunct}{\mcitedefaultseppunct}\relax
\EndOfBibitem
\bibitem[Yamamoto and Tanaka(2005)]{Yamamoto:2005}
J.~Yamamoto and H.~Tanaka, \emph{Nature Materials}, 2005, \textbf{4},
  75--80\relax
\mciteBstWouldAddEndPuncttrue
\mciteSetBstMidEndSepPunct{\mcitedefaultmidpunct}
{\mcitedefaultendpunct}{\mcitedefaultseppunct}\relax
\EndOfBibitem
\bibitem[Ho{\l}yst(1991)]{Holyst91}
R.~Ho{\l}yst, \emph{Physical Review A}, 1991, \textbf{44}, 3692--3709\relax
\mciteBstWouldAddEndPuncttrue
\mciteSetBstMidEndSepPunct{\mcitedefaultmidpunct}
{\mcitedefaultendpunct}{\mcitedefaultseppunct}\relax
\EndOfBibitem
\bibitem[Lei \emph{et~al.}(1995)Lei, Safinya, and Bruinsma]{Lei95}
N.~Lei, C.~R. Safinya and R.~F. Bruinsma, \emph{J Phys II France}, 1995,
  \textbf{5}, 1155--1163\relax
\mciteBstWouldAddEndPuncttrue
\mciteSetBstMidEndSepPunct{\mcitedefaultmidpunct}
{\mcitedefaultendpunct}{\mcitedefaultseppunct}\relax
\EndOfBibitem
\bibitem[Safran(1994)]{safran_book}
S.~A. Safran, \emph{Statistical thermodynamics of surfaces, interfaces, and
  membranes}, Addison-Wesley, Reading, Massachusetts, 1994\relax
\mciteBstWouldAddEndPuncttrue
\mciteSetBstMidEndSepPunct{\mcitedefaultmidpunct}
{\mcitedefaultendpunct}{\mcitedefaultseppunct}\relax
\EndOfBibitem
\bibitem[B\'eneut \emph{et~al.}(2008)B\'eneut, Constantin, Davidson, Dessombz,
  and Chan\'eac]{Beneut08}
K.~B\'eneut, D.~Constantin, P.~Davidson, A.~Dessombz and C.~Chan\'eac,
  \emph{Langmuir}, 2008, \textbf{24}, 8205\relax
\mciteBstWouldAddEndPuncttrue
\mciteSetBstMidEndSepPunct{\mcitedefaultmidpunct}
{\mcitedefaultendpunct}{\mcitedefaultseppunct}\relax
\EndOfBibitem
\bibitem[Helfrich(1978)]{Helfrich78}
W.~Helfrich, \emph{Z. NaturForsch.}, 1978, \textbf{A 83}, 305\relax
\mciteBstWouldAddEndPuncttrue
\mciteSetBstMidEndSepPunct{\mcitedefaultmidpunct}
{\mcitedefaultendpunct}{\mcitedefaultseppunct}\relax
\EndOfBibitem
\bibitem[Roux and Safinya(1988)]{Roux88}
D.~Roux and C.~R. Safinya, \emph{Journal de Physique}, 1988, \textbf{49},
  307--318\relax
\mciteBstWouldAddEndPuncttrue
\mciteSetBstMidEndSepPunct{\mcitedefaultmidpunct}
{\mcitedefaultendpunct}{\mcitedefaultseppunct}\relax
\EndOfBibitem
\bibitem[Jho \emph{et~al.}(2010)Jho, Kim, Safran, and Pincus]{Jho10}
Y.~S. Jho, M.~W. Kim, S.~A. Safran and P.~A. Pincus, \emph{The European
  Physical Journal E}, 2010, \textbf{31}, 207--214\relax
\mciteBstWouldAddEndPuncttrue
\mciteSetBstMidEndSepPunct{\mcitedefaultmidpunct}
{\mcitedefaultendpunct}{\mcitedefaultseppunct}\relax
\EndOfBibitem
\bibitem[Helfrich(1973)]{Helfrich73}
W.~Helfrich, \emph{Z. NaturForsch.}, 1973, \textbf{C 28}, 693\relax
\mciteBstWouldAddEndPuncttrue
\mciteSetBstMidEndSepPunct{\mcitedefaultmidpunct}
{\mcitedefaultendpunct}{\mcitedefaultseppunct}\relax
\EndOfBibitem
\bibitem[de~Gennes(1969)]{degennes69}
P.-G. de~Gennes, \emph{J. Phys. (Paris), Colloq.}, 1969, \textbf{30},
  C4--65\relax
\mciteBstWouldAddEndPuncttrue
\mciteSetBstMidEndSepPunct{\mcitedefaultmidpunct}
{\mcitedefaultendpunct}{\mcitedefaultseppunct}\relax
\EndOfBibitem
\bibitem[A.(1972)]{Caille72}
C.~A., \emph{C. R. S\'eances Acad. Sci., Ser. B}, 1972, \textbf{174}, 891\relax
\mciteBstWouldAddEndPuncttrue
\mciteSetBstMidEndSepPunct{\mcitedefaultmidpunct}
{\mcitedefaultendpunct}{\mcitedefaultseppunct}\relax
\EndOfBibitem
\bibitem[Petrache \emph{et~al.}(1998)Petrache, Gouliaev, Tristram-Nagle, Zhang,
  Suter, and Nagle]{Petrache98}
H.~I. Petrache, N.~Gouliaev, S.~Tristram-Nagle, R.~Zhang, R.~M. Suter and J.~F.
  Nagle, \emph{Physical Review E}, 1998, \textbf{57}, 7014--7024\relax
\mciteBstWouldAddEndPuncttrue
\mciteSetBstMidEndSepPunct{\mcitedefaultmidpunct}
{\mcitedefaultendpunct}{\mcitedefaultseppunct}\relax
\EndOfBibitem
\bibitem[Freyssingeas \emph{et~al.}(1996)Freyssingeas, Nallet, and
  Roux]{Freyssingeas96}
E.~Freyssingeas, F.~Nallet and D.~Roux, \emph{Langmuir}, 1996, \textbf{12},
  6028\relax
\mciteBstWouldAddEndPuncttrue
\mciteSetBstMidEndSepPunct{\mcitedefaultmidpunct}
{\mcitedefaultendpunct}{\mcitedefaultseppunct}\relax
\EndOfBibitem
\bibitem[Chaikin and Lubensky(1995)]{Chaikin_book}
P.~M. Chaikin and T.~C. Lubensky, \emph{Principles of condensed matter
  physics}, Cambridge University Press, Cambridge, US, 1995\relax
\mciteBstWouldAddEndPuncttrue
\mciteSetBstMidEndSepPunct{\mcitedefaultmidpunct}
{\mcitedefaultendpunct}{\mcitedefaultseppunct}\relax
\EndOfBibitem
\bibitem[{P. Sens} and {M.S. Turner}(1997)]{Sens97Jphys}
{P. Sens} and {M.S. Turner}, \emph{J. Phys. II France}, 1997, \textbf{7},
  1855--1870\relax
\mciteBstWouldAddEndPuncttrue
\mciteSetBstMidEndSepPunct{\mcitedefaultmidpunct}
{\mcitedefaultendpunct}{\mcitedefaultseppunct}\relax
\EndOfBibitem
\bibitem[Turner and Sens(1997)]{Turner97PRE}
M.~S. Turner and P.~Sens, \emph{Phys. Rev. E}, 1997, \textbf{55},
  R1275--R1278\relax
\mciteBstWouldAddEndPuncttrue
\mciteSetBstMidEndSepPunct{\mcitedefaultmidpunct}
{\mcitedefaultendpunct}{\mcitedefaultseppunct}\relax
\EndOfBibitem
\bibitem[Turner and Sens(1998)]{Turner98PRE}
M.~S. Turner and P.~Sens, \emph{Phys. Rev. E}, 1998, \textbf{57},
  823--828\relax
\mciteBstWouldAddEndPuncttrue
\mciteSetBstMidEndSepPunct{\mcitedefaultmidpunct}
{\mcitedefaultendpunct}{\mcitedefaultseppunct}\relax
\EndOfBibitem
\bibitem[Sens and Turner(2001)]{Sens01EPJE}
P.~Sens and M.~Turner, \emph{The European Physical Journal E}, 2001,
  \textbf{4}, 115--120\relax
\mciteBstWouldAddEndPuncttrue
\mciteSetBstMidEndSepPunct{\mcitedefaultmidpunct}
{\mcitedefaultendpunct}{\mcitedefaultseppunct}\relax
\EndOfBibitem
\bibitem[de~Gennes and Prost(1993)]{deGennes_book}
P.-G. de~Gennes and J.~Prost, \emph{The Physics of Liquid Crystals}, Oxford
  Science Publications, Oxford, 1993\relax
\mciteBstWouldAddEndPuncttrue
\mciteSetBstMidEndSepPunct{\mcitedefaultmidpunct}
{\mcitedefaultendpunct}{\mcitedefaultseppunct}\relax
\EndOfBibitem
\bibitem[Tovkach \emph{et~al.}(2013)Tovkach, Fukuda, and Lev]{Tovkach13PRE}
O.~M. Tovkach, J.-i. Fukuda and B.~I. Lev, \emph{Phys. Rev. E}, 2013,
  \textbf{88}, 052502\relax
\mciteBstWouldAddEndPuncttrue
\mciteSetBstMidEndSepPunct{\mcitedefaultmidpunct}
{\mcitedefaultendpunct}{\mcitedefaultseppunct}\relax
\EndOfBibitem
\bibitem[Santangelo and Kamien(2003)]{Santangelo03PRL}
C.~D. Santangelo and R.~D. Kamien, \emph{Phys. Rev. Lett.}, 2003, \textbf{91},
  045506\relax
\mciteBstWouldAddEndPuncttrue
\mciteSetBstMidEndSepPunct{\mcitedefaultmidpunct}
{\mcitedefaultendpunct}{\mcitedefaultseppunct}\relax
\EndOfBibitem
\bibitem[Yang \emph{et~al.}(1999)Yang, Weiss, Harroun, Heller, and
  Huang]{Yang:1999}
L.~Yang, T.~Weiss, T.~Harroun, W.~Heller and H.~Huang, \emph{Biophys. J.},
  1999, \textbf{77}, 2648--2656\relax
\mciteBstWouldAddEndPuncttrue
\mciteSetBstMidEndSepPunct{\mcitedefaultmidpunct}
{\mcitedefaultendpunct}{\mcitedefaultseppunct}\relax
\EndOfBibitem
\bibitem[Constantin(2010)]{Constantin:2010}
D.~Constantin, \emph{The Journal of Chemical Physics}, 2010, \textbf{133},
  144901\relax
\mciteBstWouldAddEndPuncttrue
\mciteSetBstMidEndSepPunct{\mcitedefaultmidpunct}
{\mcitedefaultendpunct}{\mcitedefaultseppunct}\relax
\EndOfBibitem
\bibitem[Lado(1967)]{Lado:1967}
F.~Lado, \emph{J. Chem. Phys.}, 1967, \textbf{47}, 4828\relax
\mciteBstWouldAddEndPuncttrue
\mciteSetBstMidEndSepPunct{\mcitedefaultmidpunct}
{\mcitedefaultendpunct}{\mcitedefaultseppunct}\relax
\EndOfBibitem
\bibitem[Lado(1968)]{Lado:1968}
F.~Lado, \emph{J. Chem. Phys.}, 1968, \textbf{49}, 3092\relax
\mciteBstWouldAddEndPuncttrue
\mciteSetBstMidEndSepPunct{\mcitedefaultmidpunct}
{\mcitedefaultendpunct}{\mcitedefaultseppunct}\relax
\EndOfBibitem
\bibitem[Sharma \emph{et~al.}(2010)Sharma, Aswal, and Kumaraswamy]{Sharma:2010}
K.~P. Sharma, V.~K. Aswal and G.~Kumaraswamy, \emph{The Journal of Physical
  Chemistry B}, 2010, \textbf{114}, 10986--10994\relax
\mciteBstWouldAddEndPuncttrue
\mciteSetBstMidEndSepPunct{\mcitedefaultmidpunct}
{\mcitedefaultendpunct}{\mcitedefaultseppunct}\relax
\EndOfBibitem
\bibitem[Boles \emph{et~al.}(2016)Boles, Engel, and Talapin]{Boles:2016}
M.~A. Boles, M.~Engel and Talapin, \emph{Chemical Reviews}, 2016, \textbf{116},
  11220--11289\relax
\mciteBstWouldAddEndPuncttrue
\mciteSetBstMidEndSepPunct{\mcitedefaultmidpunct}
{\mcitedefaultendpunct}{\mcitedefaultseppunct}\relax
\EndOfBibitem
\bibitem[Watson(1966)]{Watson_book}
G.~Watson, \emph{A Treatise on the Theory of Bessel Functions (2nd ed.)},
  Cambridge University Press, Cambridge, US, 1966, p. 435\relax
\mciteBstWouldAddEndPuncttrue
\mciteSetBstMidEndSepPunct{\mcitedefaultmidpunct}
{\mcitedefaultendpunct}{\mcitedefaultseppunct}\relax
\EndOfBibitem
\end{mcitethebibliography}
\end{document}